\documentclass{edp-conf} \usepackage{graphicx} \usepackage{epsfig}

\newcommand{\Mpc}{\,{\rm Mpc}}
\newcommand\ltsima{$\; \buildrel < \over \sim \;$}
\newcommand\simlt{\lower.5ex\hbox{\ltsima}}
\newcommand\gtsima{$\; \buildrel > \over \sim \;$}
\newcommand{\gev}{\,{\rm GeV}}
\newcommand\simgt{\lower.5ex\hbox{\gtsima}}

\TitreGlobal{SF2A 2005}

\begin{document}
\title{Charting the New Frontier of the Cosmic Microwave Background Polarization}

\author{F.~R.~Bouchet} \address{IAP, 98bis Bd. Arago, 75014 PARIS, France}
\author{A.~Beno{\^\i}t \& Ph. Camus} \address{CRTBT, 25 av. des Martyrs, F-38042 Grenoble,
France}
\author{F.~X.~D{\'e}sert} \address{LAOG, BP 53, 414 rue de la piscine, F-38041 Grenoble,
France}
\author{M.~Piat} \address{APC-Coll{\`e} ge de France, 11 place M. Berthelot, F-75231
PARIS}
\author{N.~Ponthieu} \address{IAS, Bat. 121, Universit{\'e} Paris-Sud, F-91405 ORSAY}

\runningtitle{Charting the polarization frontier with SAMPAN}
\setcounter{page}{1}

\index{Bouchet, F.~R.}  \index{Beno{\^\i} t, A.}  \index{Camus, Ph.}  \index{D{\'e} sert,
F.~X.}  \index{Piat, M.}  \index{Ponthieu, N.}

\maketitle

\begin{abstract} 
The anisotropies of the cosmic microwave background are a gold mine for cosmology and
fundamental physics. ESA's Planck satellite should soon extract all information from
the temperature vein but will be limited concerning the measurement of the degree of
polarization of the anisotropies. This polarization information allows new independent
tests of the standard cosmological paradigm, improves knowledge of cosmological
parameters and last but not least is the best window available for constraining the
physics of the very early universe, particularly the expected background of primordial
gravitational waves. But exploiting this vein will be a challenge, since the
sensitivity required is {\em at least} 10 times better than what Planck might achieve
at best, with the necessary matching level of control of all systematics effects, both
instrumental and astrophysical (foregrounds). We here recall the cosmological context
and the case for CMB polarization studies. We also briefly introduce the SAMPAN
project, a design study at CNES that aims at detecting the primoridal gravitational
wave background for a tensor to scalar ratio $T/S$ as small as $10^{-3}$.
\end{abstract}

\begin{figure}[ht]
\centering \includegraphics[width=9cm]{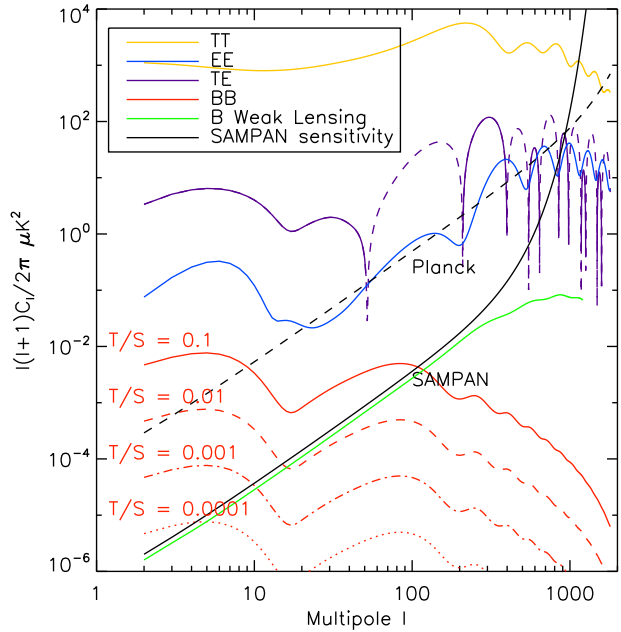}
\caption{Angular power spectra of the temperature and polarization anisotropies of the
CMB for the $\Lambda$CDM model that best fits the WMAP first year
data. The $B$ mode spectrum is presented for different
Tensor--to--Scalar ratios $T/S = 0.1$ to 0.0001, i.e. for an inflation
energy scale $E_{infl}\simeq 2.2\,10^{16},\;1.2\,10^{16},\;6.8
\,10^{15},\;3.8\,10^{15}$~GeV.  The $E$ modes transformed into $B$
modes by weak--lensing are in green. The SAMPAN target sensitivity
(5$\mu K$.arcmin) is in solid black and can be compared with that of
Planck in black dashes. SAMPAN should be able to detect the
primordial $B$ mode polarization down to $T/S \simeq 0.001$ on very
large angular scales.
}
\label{fig:cl}
\end{figure}

\paragraph{The Cosmological paradigm.}
The spatial distribution of galaxies revealed the existence of large scale structures
in the Universe, clusters of size $\sim 5\,\Mpc$, filaments connecting them, and voids
of size $\sim 50\,\Mpc$. Their existence and statistical properties can be accounted
for by the development of primordial fluctuations by gravitational instability. The
current paradigm is that these fluctuations were generated in the very early Universe,
probably during an inflationary period, that they evolved linearly during a long
period, and more recently reached density contrasts high enough to form bound objects.

The anisotropies of the CMB are the imprint of these fluctuations as they
were\footnote{But for a small correction due to the photons propagation through the
developing large scale structures.} when photons last interacted with baryonic matter,
when the Universe became neutral at a redshift of 1100 (the Universe was then $\sim
370\, 000$ years old). Since at that time the fluctuations are still tiny, all the
evolution till then is linear and can be computed quite accurately. Despite some
degeneracies, the value of most parameters can in principle be inferred from
sufficiently accurate measurement of the angular power spectra $C_\ell$ of the CMB
temperature and polarization anisotropies. The CMB polarization fluctuations are due
to the last Thomson scatterings of the photons on free electrons that see an
inhomogeneous (quadrupolar) incoming radiation field.

\paragraph{The CMB polarization.}
The main source of these quadrupoles is the very same density (and associated
velocity) perturbations that source the temperature anisotropies. This particular
state of polarization ($E$) is therefore correlated to temperature. An accurate
measure of $E$ enables to break degeneracies between the effects of some parameters,
for instance that between the reionization optical depth and the scalar perturbation
spectral index. The $E$ mode polarization has already been detected
(\cite{kovac,redhead,barkats,boo05}, see Fig.~\ref{fig:cl}) and should be further
characterized by forthcoming experiments (for a review, see \cite{ponthieu05}). A
second source of polarization of the CMB are tensorial perturbations such as those
from a background of Gravitational Waves generated during Inflation (hereafter
IGW). The induced polarization pattern is of odd parity ($E$ was of even parity
because density fluctuations are scalar) and has been named $B$ in analogy with
electromagnetic fields (\cite{seljak}). No other primordial source than the IGW could
generate $B$ type polarization. Its detection would therefore be a signature of the
IGW background. A measurement of the amplitude of the $B$ mode power spectrum gives a direct
estimation of the energy scale of inflation. Interestingly enough, CMB polarization
experiments can probe $E_{inf}$ at the typical GUT scale, which means that they have a
strong constraining power on the relation between inflation and GUT, which has
implications not only in Cosmology but also in High Energy Physics. This is way beyond
the capabilities of next generation gravitational interferometers. A challenge that a
CMB polarization experiment faces is the abitility to remove foregrounds emission (see
below). Another challenge would be to actually subtract the part of $E$ mode power
which is transformed into $B$ modes though their (weak) lensing by large scale
structures~(\cite{knox}). While possible in principle, this calls for very high
sensitivity and angular resolution, but it would then allow the detection of weaker
B-contributions.

The detection of the $B$ mode polarization requires large improvements on instrumental
sensitivity, systematic effects and foregrounds control. Indeed, the desired
sensitivity is 10 to $30$ times better than Planck should do. Since detectors are already nearly
photon noise limited, the only solution is to increase their number in the focal
planes and to go up to several 10,000s which has dramatic implications in terms of
electronics, cryogenics and telemetry. This is the goal of the ongoing bolometer array
developpements.

\paragraph{The SAMPAN project.}
One may then ask what would be the best experimental design to detect the $B$ modes?
The gravity wave signal is on large angular scales ($\theta > 1$~deg). The resolution
is therefore not an issue in itself and a subdegree beam would do (if one does not
attempt to clean the $B$ map from its lensed $E$ mode contribution). On the other hand,
the coverage of the sky should be as complete as possible, since 1) $B$ modes are
largest on large scales, 2) increasing coverage decreases cosmic variance 3) missing
information impinges the decomposition of polarisation into separate $E$ and $B$ modes.

The polarization of foregrounds is a serious issue for CMB studies, especially for
polarization. Recent results from Archeops (\cite{arch_polar_1,arch_polar_2}) have
shown that the thermal emission from dust is significantly polarized on large angular
scales. To control this radiation field, several frequency channels are required. The
100, 143, 217 and 353~GHz bands, advocated for Planck, optimize the sensitivity to the
CMB and should enable to remove the dust contribution at the best possible level.

{\bf SAMPAN} (\emph{SAtellite to Measure the Polarized ANisotropies}) is a
mini-satellite project at CNES to be launched after 2012 to an L2 orbit. It is
currently in ``phase 0'' of preliminary design study. The goal is to have $20\;000$
polarization sensitive bolometers at 100~mK, divided in the four frequency bands
mentionned above, in order to reach a target (combined) CMB sensitivity of $5\;\mu 
K$.arcmin. This would enable the detection of the primordial $B$ mode if the Tensor to
Scalar ratio is $T/S \simgt 0.001$, i.e. the energy scale of inflation $E \simgt 6 \times 
10^{15}\gev$.  This is close to the limit anticipated for foregrounds removal
(\cite{tucci}).

In order to have as symmetrical a system as possible, we are studying a refractive
telecentric system of at least $20$~arcmin resolution FWHM at 217~GHz. 
The main originality of SAMPAN is its scanning strategy dedicated to have redundancies
at many different timescales. First the polarization measurement can be checked by
spinning the whole satellite around its main optical axis in a period between 10 and
20 seconds.  A precession/nutation motion of the satellite then allows connecting
large angular scales within a short time span (40~deg.  separation on the sky in a few
minutes) and to cover half of the sky in a time scale of few days.

\paragraph{Conclusions:}
The CMB remains unique in tightening together so many fundamental elements:
fundamental physical laws, cosmography and cosmogony, i.e the cosmological paradigm
for the Universe content, evolution, structuring, and its parameters. Building such a
polarization mission will surely be challenging, but in proportion to the potential
pay-offs in addressing such topics as tighter tests of the cosmological paradigm, the
primordial gravitational wave background and the energy scale of inflation, or
determining neutrinos masses.

\end{document}